\documentclass[a4paper]{VisionStyle}
\usepackage{epsfig}

\begin{document}

\title{The $log$ $N$ - $log$ $S$ and the Broadband Properties
of the Sources in the HELLAS2XMM Survey}

\author{A.\,Baldi\inst{1} \and S.\,Molendi\inst{1} \and 
  A.\,Comastri\inst{2} \and F.\,Fiore\inst{3} \and G.\,Matt\inst{4} \and 
  C.\,Vignali\inst{5} } 

\institute{
  Istituto di Fisica Cosmica - CNR, via Bassini 15, I-20133 Milano, Italy 
\and 
  Osservatorio Astronomico di Bologna, via Ranzani 1, I-40127 Bologna, Italy
\and 
  Osservatorio Astronomico di Roma, via Frascati 33, I-00040 Monteporzio, Italy  
\and
  Dipartimento di Fisica - Universit\'a di Roma Tre, via della Vasca 
Navale 84, I-00146 Roma, Italy  
\and
  Dept. of Astronomy and Astrophysics - The Pennsylvania
State University, 525 Davey Lab, University Park, PA 16802 USA }

\maketitle 

\begin{abstract}

We present the first results from an {\it XMM-Newton} serendipitous medium-deep 
survey, which covers nearly three square degrees. We show the $log$ $N$ - $log$ $S$ 
distributions for the 0.5-2, 2-10 and 5-10 keV bands, which are found to be in 
good agreement with previous determinations and with the predictions of XRB 
models. In the soft band we detect a break at fluxes around $5\times10^{-15}$ cgs.
In the harder bands, we fill in the gap at intermediate fluxes between deeper 
{\it Chandra} and {\it XMM-Newton} observations and shallower {\it BeppoSAX} 
and {\it ASCA} surveys.
Moreover, we present an analysis of the broad-band properties of the sources.

\end{abstract}

\section{Introduction}
While in the soft band (0.5-2 keV) {\it ROSAT} (\cite{abaldi-F:has98}) and 
especially 
{\it Chandra} (\cite{abaldi-F:ros01}) has resolved almost all the XRB, in 
the hard band (2-10 keV) the XRB has been resolved
at a 25\%-30\% level with {\it BeppoSAX} and {\it ASCA} surveys 
(\cite{abaldi-F:cag98}; \cite{abaldi-F:gio00}) and 
recently at a 90\% with {\it Chandra} (\cite{abaldi-F:ros01}). Moreover, 
in the very hard band (5-10 keV) the fraction resolved
by {\it BeppoSAX} is around 30\% (\cite{abaldi-F:fio99}) and recently in the 
{\it XMM-Newton} Lockman Hole deep pointing a 60\% is reached 
(\cite{abaldi-F:has01}).\\
The optical counterparts of the objects making the XRB are predominantly 
Active Galactic Nuclei (AGN).
In the soft band the predominant fraction is made by unabsorbed AGN, with a 
small fraction of absorbed AGN (\cite{abaldi-F:sch98}).
The fraction of absorbed type-2 AGN rises if we consider the spectroscopic
identifications of hard X-ray sources in {\it BeppoSAX}, {\it ASCA} and 
{\it Chandra} surveys
(\cite{abaldi-F:fio01}; \cite{abaldi-F:del00}; \cite{abaldi-F:toz01}).\\
The X-ray and optical observations are consistent with current 
XRB synthesis models (\cite{abaldi-F:com95}; \cite{abaldi-F:gil01}),
which explain the hard XRB spectrum with an appropriate mixture of absorbed 
and unabsorbed AGN, by introducing the corresponding luminosity function 
and cosmological evolution.
However, these models require the presence of a 
significant population of type-2 QSOs (\cite{abaldi-F:nor01}), not yet 
detected in sufficient quantities.
Type-2 QSOs are rare (so far, only a few are known), 
luminous and hard (heavily absorbed in the soft band). A good way of 
finding them is to perform surveys in the hard X-ray bands, covering 
large solid angles.
The large throughput and effective area, particularly in the harder bands, 
make {\it XMM-Newton} currently the best satellite to perform hard X-ray
surveys.\\
In this poster contribution we present results from the HELLAS2XMM survey
(\cite{abaldi-F:bal02}), one of its main goals is to constrain the 
contribution of absorbed AGN to the XRB.   

\section{Data preparation and cleaning}
\label{abaldi-F_sec:dataprep}

We use the XMM-SAS analysis software tasks $epproc$ and
$emproc$ to linearize the event files.
Before processing, datasets are corrected for the attitude of the 
satellite in order to have absolute positions in the sky.
The Attitude History File(AHF) coordinates are 
given to the SAS task $odffix$ which performs the correction.\\
The event files produced by $epproc$ and $emproc$ are cleaned from:
\begin{itemize}
\item hot pixels; with a procedure (developed at IFC/CNR-Milan by A. De Luca) 
which uses cosmic ray IRAF tasks to localize the pixels to be rejected in each 
CCD and XMM-SAS task $evselect$ for removing them from the event files;
\item soft proton flares; analysing the light curves at energies 
greater than 10keV and setting a threshold for good time intervals of 0.15 
cts/s for MOS units and of 0.35 cts/s for pn unit.
\end{itemize}
A complete set of MOS1, MOS2 and pn 600x600 pixel images (1 pixel = 
4.35 arcsec) is generated using XMM-SAS task $evselect$ in the 0.5-2,
2-10, 2-4.5, 4.5-10 and 0.5-10 keV bands.
MOS and pn images are merged together in order to increase the signal-to-noise
ratio of the sources and go deeper.\\
A corresponding set of exposure maps is generated to account for 
spatial quantum efficiency, mirror vignetting and field of view of each
instrument, running XMM-SAS task $eexpmap$.
The so-created exposure maps are not completely satisfactory, since the
evaluation of quantum efficiency, filter transmission, and vignetting is
performed assuming an event energy which corresponds to the mean of
the energy boundaries. In the 2-10 keV band, this may lead to inaccuracies
in the estimate of these key quantities, thus we create the 2-10 keV 
band exposure map as a weighted mean of the 2-4.5 keV and the 4.5-10 keV
exposure maps (assuming an underlying power-law spectral model with 
$\Gamma=1.7$).\\
\begin{figure}[ht]
  \begin{center}
    \epsfig{file=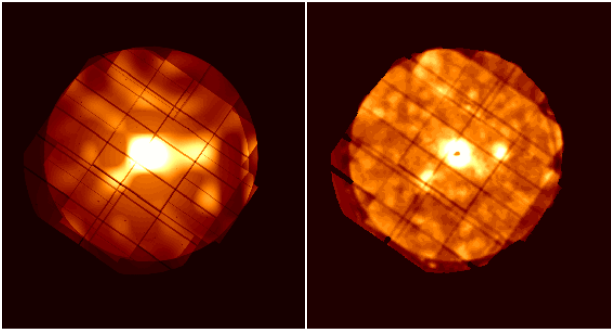, width=4.5cm, angle=-90}
  \end{center}
\caption{Left: {\it XMM-SAS} standard background map. 
Right: corrected background map.}
\label{abaldi-F_fig:bkgmap}
\end{figure}
XMM-SAS task $esplinemap$ creates a background map by
removing all the sources above a fixed maximum likelihood
threshold and fitting the remaining with a cubic spline.
Even using the maximum number of spline nodes (20), the fit is not
enough flexible to take into account the local variations of the
background. We correct the background map produced by XMM-SAS, pixel
by pixel, comparing the mean value in it with that of the so-called
cheesed image, within a radius of $3\cdot r_{0.68}$ ($r_{0.68}$ 
= radius corresponding to an encircled energy fraction (EEF) of 0.68).
A comparison between XMM-SAS standard background map and our corrected
map is shown in Figure~\ref{abaldi-F_fig:bkgmap}.\\

\section{Source detection and characterization}
\label{abaldi-F_sec:srcdet}

A preliminary detection run, using XMM-SAS $eboxdetect$, is
performed in each energy band, in order to create a list of candidate sources.
Each source is characterized using a radius corresponding to an EEF
of $\alpha$=0.68.\\
The source counts $S$ and their error $\sigma_S$ are determined as
$$
S=\frac{cts_{src}-cts_{bkg}}{\alpha}
$$
$$
\sigma_S=\frac{1+\sqrt{cts_{src}+0.75}}{\alpha}
$$
The countrate is $cr=\frac{S}{T_{tot}}$ where $T_{tot}$ is the sum of MOS1, MOS2
and pn exposure times. The corresponding flux is $F_x=cf\cdot cr$ where $cf$
is calculated from the
$$
\frac{T_{tot}}{cf}=\frac{T_{MOS1}}{cf_{MOS1}}+\frac{T_{MOS2}}{cf_{MOS2}}+\frac{T_{pn}}{cf_{pn}}
$$
Moreover we compute $p$, the poissonian probability that counts originate from
a background fluctuation, from the
$$
\sum_{n=cts_{src}}^\infty e^{-cts_{bkg}}\frac{cts_{bkg}^n}{n!}>p
$$
and choose a threshold of $p=2\times10^{-4}$ to decide whether 
to accept or not a detected source.

\section{The survey}
\label{abaldi-F_sec:survey}

\begin{figure}[ht]
  \begin{center}
    \epsfig{file=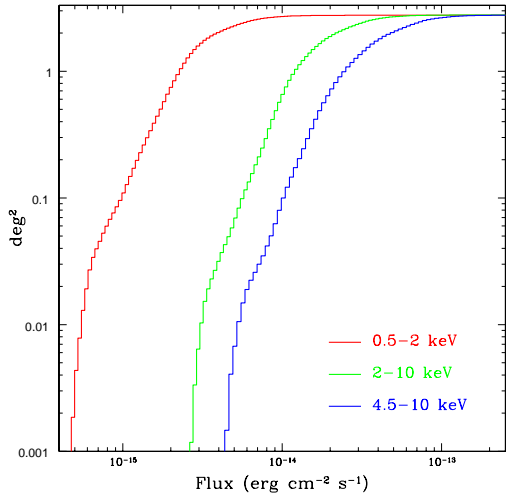, width=7cm}
  \end{center}
\caption{The total sky coverage of the survey in the 0.5-2 keV (red), 2-10 keV
(green) and 4.5-10 keV band (blue).}
\label{abaldi-F_fig:skycov}
\end{figure}

The HELLAS2XMM survey (\cite{abaldi-F:bal02}) currently uses the 15 {\it 
XMM-Newton} calibration and 
performance verification phase fields shown in table~\ref{abaldi-F_tab:15fields}.
All the fields are at high galactic latitude ($|bII|$ $>$ 27$^o$), have low 
galactic $N_H$ (a few $10^{20}$ cm$^{-2}$) and at least 15 ksec of good 
observing time.
\begin{table}[bht]
  \caption{The HELLAS2XMM survey sample.}
  \label{abaldi-F_tab:15fields}
  \begin{center}
    \leavevmode
    \footnotesize
    \begin{tabular}[h]{lrr}
      \hline \\[-5pt]
Target&N$_H$ (cm$^{-2}$)&$bII$($^o$)\\[+5pt]
      \hline \\[-5pt]
PKS0537-286&$2.1\cdot 10^{20}$&-27.3\\
PKS0312-770&$8\cdot 10^{20}$&-37.6\\
MS0737.9+7441&$3.5\cdot 10^{20}$&29.6\\
Lockman Hole&$5.6\cdot 10^{19}$&53.1\\
Mkn 205&$3\cdot 10^{20}$&41.7\\
BPM 16274&$3.2\cdot 10^{20}$&-65.0\\
MS1229.2+6430&$2\cdot 10^{20}$&52.8\\
PKS0558-504&$4.5\cdot 10^{20}$&-28.6\\
Mkn 421&$7\cdot 10^{19}$&65.0\\
Abell 2690&$1.9\cdot 10^{20}$&-78.4\\
G158-100&$2.5\cdot 10^{20}$&-74.5\\
GD153&$2.4\cdot 10^{20}$&84.7\\
IRAS13349+2438&$1.2\cdot 10^{20}$&60.6\\
Abell 1835&$2.3\cdot 10^{20}$&60.6\\
Mkn 509&$4.1\cdot 10^{20}$&-29.9\\
      \hline \\
      \end{tabular}
  \end{center}
\end{table}
The sky coverage of the sample has been computed using the exposure
maps of each instrument, the background map of the merged image and a model for 
the PSF. We adopt the off-axis angle dependent
PSF model implemented in {\it XMM-SAS} $eboxdetect$ task.\\ 
At each 
image pixel $(x,y)$ we evaluate,
within a radius $r_{0.68}$, the total background counts (from
the background map). From these we calculate 
the minimum total counts (source + background) necessary for a source 
to be detected at a probability $p=2\times10^{-4}$ (defined in 
Section~\ref{abaldi-F_sec:srcdet}).
The mean exposure times for {\it MOS1}, {\it MOS2} and {\it pn}, evaluated 
from the exposure maps within $r_{0.68}$, are used to compute the count rate 
$cr$. From the
count rate-to-flux conversion factor $cf$ (computed as in 
Section~\ref{abaldi-F_sec:srcdet}) we build a flux limit map and straightforwardly
calculate the sky coverage of a single field.\\
Summing the contribution from all fields
we obtain the total sky coverage of the survey, which is plotted in 
Figure~\ref{abaldi-F_fig:skycov}, in three different energy bands.\\

\section{The $log$ $N$ - $log$ $S$ Relation}

\begin{figure}[ht]
  \begin{center}
    \epsfig{file=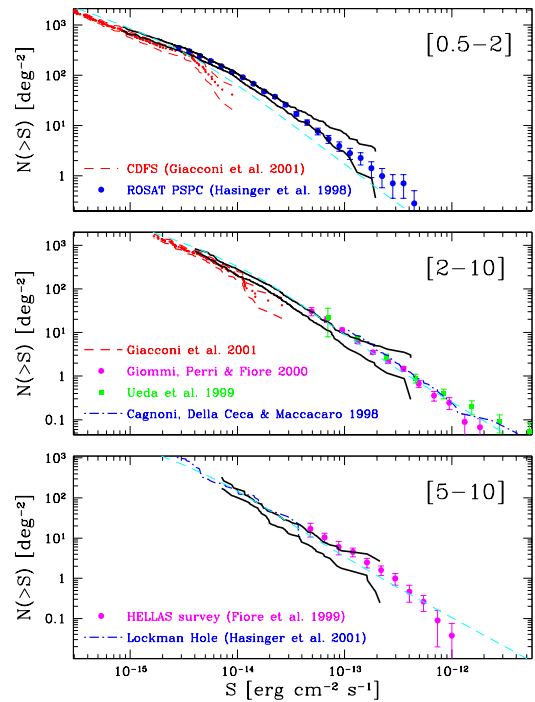, width=7cm}
  \end{center}
\caption{The cumulative $log$ $N$ - $log$ $S$ in the 0.5-2 keV (top), 
2-10 keV (center) and 5-10 keV band (bottom). In all diagrams
the black thick solid lines are the upper and lower limits of our Log(N)-Log(S).
The dashed cyan lines are the predictions of the XRB synthesis models from
Comastri et al.(2001).}
\label{abaldi-F_fig:lognlogs}
\end{figure}

The $log$ $N$ - $log$ $S$ distributions are plotted in 
Figure~\ref{abaldi-F_fig:lognlogs} and contain 1022, 495 and 100 sources,
for the 0.5-2 keV, 2-10 keV and 5-10 keV band, respectively.\\
In the 0.5-2 keV band the distribution shows a flattening around 
$5\times10^{-15}$ erg cm$^{-2}$ s$^{-1}$, similarly to {\it ROSAT} data 
(\cite{abaldi-F:has98}) although with a flatter differential slope index at 
faint fluxes. We are also in good agreement with {\it Chandra} Deep Field South
data (\cite{abaldi-F:gia01}).
In the 2-10 keV band we find that the distribution is significantly 
sub-euclidean, in contrast to {\it BeppoSAX} and {\it ASCA} findings
(\cite{abaldi-F:gio00}; \cite{abaldi-F:cag98}; \cite{abaldi-F:ued99}), 
indicating that the $log$ $N$ - $log$ $S$ flattens at fainter fluxes. Anyhow 
it represents a link between {\it Chandra} (\cite{abaldi-F:gia01}) and {\it 
BeppoSAX-ASCA} observation, sampling an intermediate flux range. 
The 5-10 keV 
$log$ $N$ - $log$ $S$ is consistent with an euclidean slope and samples an 
intermediate flux range between {\it XMM-Newton} deeper observations 
(\cite{abaldi-F:has01}) and {\it BeppoSAX} shallower HELLAS survey
(\cite{abaldi-F:fio99}).

\section{Hardness Ratio Analysis}
We divided the sample of sources detected both in 0.5-2 keV band and in 2-4.5
keV band in two subsamples containing the brighter 
($F_{0.5-2keV}>2\times10^{-14}$ erg cm$^{-2}$ s$^{-1}$) and the fainter
($F_{0.5-2keV}\le2\times10^{-14}$ erg cm$^{-2}$ s$^{-1}$) sources, respectively
and we compute for them the hardness ratio:
$$
HR_1 = \frac{cr_{2-4.5}-cr_{0.5-2}}{cr_{2-4.5}+cr_{0.5-2}}
$$
\begin{figure}[ht]
  \begin{center}
    \epsfig{file=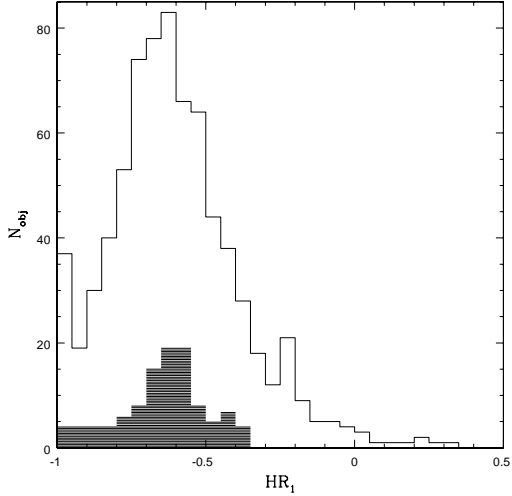, width=7cm}
  \end{center}
\caption{The distribution of hardness ratio $HR_1$. {\it Void histogram}: 
sources with $F_{0.5-2keV}\le2\times10^{-14}$ erg cm$^{-2}$ s$^{-1}$. 
{\it Shaded histogram}: sources with $F_{0.5-2keV}>2\times10^{-14}$ erg 
cm$^{-2}$ s$^{-1}$.}
\label{abaldi-F_fig:hr1}
\end{figure}
\begin{table}[bht]
  \caption{$HR1$ distribution for the faint and the bright sample.}
  \label{abaldi-F_tab:hr1distrib}
  \begin{center}
    \leavevmode
    \footnotesize
    \begin{tabular}[h]{lcc}
      \hline \\[-5pt]
&N($HR_1<-0.35$)&N($HR_1\ge-0.35$)\\[+5pt]
      \hline \\[-5pt]
Faint sample&654&84\\
Bright sample&107&0\\
      \hline \\
      \end{tabular}
  \end{center}
\end{table}
As shown in Figure~\ref{abaldi-F_fig:hr1} and in Table~\ref{abaldi-F_tab:hr1distrib},
the faint sample shows a tail of hard sources which 
is not present in the bright sample. The probability of having 84 sources with 
$HR_1\ge-0.35$ in the faint sample and no sources with $HR_1\ge-0.35$ in the 
bright sample is $\sim10^{-6}$, so the progressive hardening of the sources
towards fainter fluxes seems to be highly significant.\\
\begin{figure}[ht]
  \begin{center}
    \epsfig{file=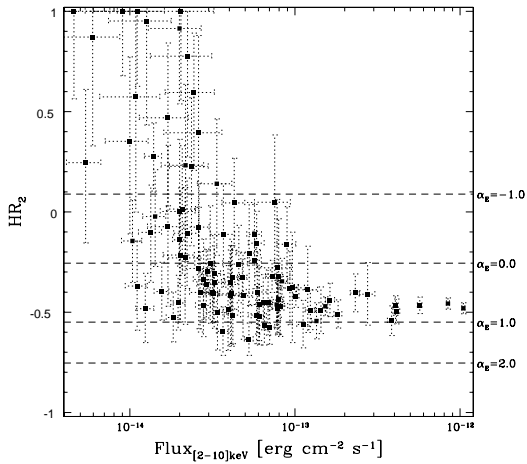, width=7cm}
  \end{center}
\caption{2-10 keV flux vs. $HR_2$ for the sources detected in the 4.5-10 keV
band. The dashed lines indicate the hardness ratios of the sources according 
to a power-law model (galactic absorption) with spectral index $\alpha_E$.}
\label{abaldi-F_fig:hr2}
\end{figure}
A further analysis of the hardness ratio has been carried out on the 
4.5-10 keV sample.
In Figure~\ref{abaldi-F_fig:hr2} the relation between the 2-10 keV flux and the
hardness ratio 
$$
HR_2  = \frac{cr_{4.5-10}-cr_{2-4.5}}{cr_{4.5-10}+cr_{2-4.5}}
$$
for sources detected in the 4.5-10 keV band is plotted. The diagram shows a 
progressive hardening of the sources when going towards fainter fluxes. While
the lack of sources in the bottom left of the figure is due to a selection
effect (because we are more sensitive in the 2-4.5 keV than in the 4.5-10 keV
band), there is no apparent reason for having no hard sources at fluxes
greater than a few $10^{-14}$ erg cm$^{-2}$ s$^{-1}$.\\
The population of harder sources we detect, probably consists of AGN having 
substantial absorbing column densities ($N_H>10^{22}$ cm$^{-2}$).

\section{Summary}

We are carrying out a serendipitous {\it XMM-Newton} survey. We currently cover 
nearly three square degrees in 15 fields observed during satellite
calibration and performance verification phase. This is, 
to date, the {\it XMM-Newton} survey with the largest solid angle.\\
The main results can be summarized as follows:
\begin{itemize}
\item The $log$ $N$-$log$ $S$ relations in the 0.5-2 keV, 2-10 keV and 
5-10 keV band are in agreement with previous determinations;
\item in the hard bands we sample an intermediate flux range: deeper 
than {\it ASCA} and {\it BeppoSAX} and shallower than {\it Chandra} and 
{\it XMM-Newton} deep surveys
\item We find an evidence for hard sources emerging below 0.5-2 keV
fluxes of $2\times10^{-14}$ erg cm$^{-2}$ s$^{-1}$ and 2-10 keV
fluxes of 10$^{-13}$ erg cm$^{-2}$ s$^{-1}$.
\end{itemize}

\begin{acknowledgements}

We thank A. De Luca for developing the hot pixel cleaning algorithm. 
We are also grateful to G. Zamorani, G. C. Perola and all members of the 
HELLAS2XMM team for useful discussions. AB and SM acknowledge partial financial
support by ASI I/R/190/00 contract.

\end{acknowledgements}

\end{document}